\documentclass[prl,reprint,superscriptaddress,twocolumn]{revtex4}

\newcommand{\ket}[1]{|#1\rangle}

\usepackage{amsmath,amssymb,amsthm,amsfonts,mathptmx}

\usepackage{algorithm}

\begin{document}
\title{Quantum Blockchain Miners Provide Massive Energy Savings}
\author{Joseph Kearney}
\affiliation{School of Computing, University of Kent, Canterbury, Kent CT2 7NF United Kingdom.}
\author{Carlos A Perez-Delgado}
\email{c.perez@kent.ac.uk}
\affiliation{School of Computing, University of Kent, Canterbury, Kent CT2 7NF United Kingdom.}

\begin{abstract}
Blockchain-based cryptocurrencies have become an extremely important, highly-used, technology. A major criticism of cryptocurrencies---however---is their energy consumption.
In May 2022 Bitcoin alone was reported to be consuming 150 terawatt-hours of electricity annually---more than many entire countries. Hence, any meaningful efficiency increase in this process would have a tremendous positive impact. Meanwhile, practical applications of quantum information technologies, and in particular of near-term quantum computers (NISQ) continue to be an important research question.
Here, we study the efficiency benefits of moving cryptocurrency mining from current ASIC-based miners to \emph{quantum}, and in particular NISQ, miners. While the time-efficiency benefits of quantum technologies is extremely well-studied, here we focus on energy savings. We show that the transition to quantum-based mining could incur an energy saving---by relatively conservative estimates---of about roughly 126.7TWH, or put differently  the total energy consumption of Sweden in 2020.
\end{abstract}

\maketitle

\section{Introduction}\label{sec:intro}

Blockchain technologies  have exploded into a thriving industry, with thousands of protocols and projects. The industry is based on the principle of decentralised transfer of data and digital assets, secured and intrinsically coupled with its cryptography. One of the major applications of blockchain technologies is on \emph{cryptocurrencies}\cite{jalal2021bibliometric}. Cryptocurrencies aim to replicate many of the features of \emph{fiat} currency, while adding some features specific to blockchain technology, such as decentralisation, anonymity, non-repudiation, and high availability backed up by state-of-the-art cryptographic technology. While the cryptocurrency market is highly volatile, it has largely trended upward, since its introduction in 2008. In April 2021, Bitcoin alone was worth over one trillion US dollars ($10^{12}$USD).

One of the underlying protocols that make a blockchain work is that of \emph{consensus}. A consensus protocol allows hundreds or even millions of blockchain users to agree on how to add a new \emph{block}---essentially a chunk of data representing transactions---to the blockchain. By far, the most widely used consensus protocol in cryptocurrencies today is Proof-of-Work (PoW)\cite{bach2018comparative}.

In a PoW-based blockchain, any user wishing to add a block to the chain must first solve a computationally difficult numerical problem (work) and provide the solution (proof) to the entire network. This process is called \emph{mining}. The term \emph{miner} is used to denote both the agent/user, and the hardware, involved in the mining.

One major criticism of PoW based blockchain technologies is their excessive energy consumption \cite{sutherland2019blockchain, schinckus2021proof}. This high energy usage is currently having a notable impact on global carbon emissions \cite{stoll2019carbon}.  

The extent of this energy consumption is well documented \cite{zhang2020evaluation, taylor2018analysis, fiat2019energy}. 
At the time of writing Cambridge Bitcoin Electricity Consumption Index (CBECI) \cite{ccaf.io_2022} calculates Bitcoins energy consumption to be 0.57\% of the worlds total produced energy. This is 126.7 TWh per year, which is more than Norway (124.3 TWh per year). This is the energy cost of keeping just \emph{one}   of many cryptocurrencies  running (Bitcoin).

Given the current energy costs, and general upward trend, the need for more energy efficient cryptocurrency mining is clear. Enter quantum technologies.

Quantum computers, and their theoretical advantages are very well studied\cite{kearney2021vulnerability, daley2022practical, ekeraa2017quantum, riste2017demonstration}. One advantage is in the time efficiency  of quantum computers, over their classical counterparts, for various computationally difficult numerical problems. This general advantage applies to the particular case of PoW mining\cite{bard2022quantum}.

In this paper, however, we are interested in the energy efficiency advantages. Quantum computers derive an energy efficiency advantage from two different sources. The first is the time-efficiency advantage already mentioned: if a device can perform the same task, in a shorter time-period, this will often convey an energy advantage as well.

The second is slightly more subtle. Quantum computation requires (mostly) \emph{unitary} evolution. Unitary evolution, by its nature, is reversible, and hence, to a large extent, energy neutral. At it's theoretical limit, a quantum cryptocurrency miner could potentially expend little to no energy. 

Achieving this \emph{Landaeur}-limited efficiency is unrealistic. In a real world setting, implementing and running a quantum miner would entail unavoidable inefficiencies. It would be very difficult to account for all engineering decisions and inefficiencies, individually. 

However, we can make a meaningful prediction of the energy-savings achieved by quantum miners if we make one \emph{fairly reasonable} assumption: that the \emph{inefficiency} of a quantum miner---that is, the amount of energy consumed, and heat expelled, over its theoretical optimum---is within an order of magnitude of a \emph{classical miner's inefficiency.}

The above assumption is, on its face, a reasonable one, and allows us to predict an actual energy consumption number for quantum miners. That said, we can take an extra step, and ask
 just how much more inefficient a quantum miner would have to be, before it lost its advantage over classical mining systems altogether. 

Finally, we don't just study the potential of quantum miners, we in particular study the potential of Noisy Intermediate-Scale Quantum\cite{preskill2018quantum} (NISQ) based cryptominers. This is notable for two reasons. First, being able to implement cryptominers on NISQ hardware, rather than having to wait for full-blown scalable fault-tolerant quantum computers, would greatly accelerate the possible adoption of quantum miners. The second is that since even before the coinage of the term, the search for practical applications of NISQ technology has been an important ongoing one.

In the next section we summarise our findings, and in the section afterwards we describe in detail how we derived them.. While the  analysis that follows works for any PoW-based cryptocurrency, for the sake of exposition we will focus on the Bitcoin network.

\section{Results}\label{sec:RES}

We consider three different quantum mining systems, and compare them to a classical miner. 

The classical miner we are comparing our quantum miners to is a current state-of-the-art ASIC cryptocurrency miner: an Antminer S19 XP, ASIC miner.

We consider two different miner implementations using fully scalable, fault-tolerant quantum computers. The first is on a system using a two-layer Shor code that can tolerate a (large) $0.1$ physical qubit error rate, while still successfully completing its computation with probability greater than $1 - 1.23 \times 10^{-6}$.

The second uses a single layer Shor code, hence only being able to tolerate a physical qubit error rate of $0.0001$ to achieve the same success probability.

Finally, we also consider a non-ECC NISQ computer. For this computer, we do not consider a target success probability, as this would depend greatly on the physical error-rate of the NISQ architecture that may vary widely---rather we focus solely on the energy costs. The assumption here is that, given the nature of cryptomining in general, a relatively large failure-rate that would not be workable in other contexts, may very well be considered acceptable, and even profitable. More on this in the Conclusions.

Our results are summarised in Table \ref{tab:TAB2}.  The first row of Table \ref{tab:TAB1} (classical miner) is directly calculated. The costs for the other three rows (three different quantum miners), are all estimated; see the Methodology Section for details.

\begin{table}[ht] 
\centering
\begin{tabular}{|p{2.2cm}|p{2.0cm}|p{2.0cm}|p{1.9cm}|}
\hline
\textbf{Infrastructure} &
  \textbf{Landauer Theoretical Minimum (J)} &
  \textbf{Ratio (1:379) (J)} &
  \textbf{Ratio (1:1706) (J)} \\ \hline
Classical                      & 1.324  & 502  & 2258.69  \\ \hline
Non-ECC NISQ Miner         & $1.43 \times10^{-18}$  & $5.43 \times10^{-16}$  & $1.43 \times10^{-15}$ \\ \hline
1 Layer ECC Quantum Miner & $3.75 \times10^{-10}$ & $1.42 \times10^{-7}$ & $6.4 \times10^{-7}$ \\ \hline
2 Layer ECC Quantum Miner & $4.5 \times10^{-9}$ & $1.70 \times10^{-6}$ & $7.68 \times10^{-6}$ \\ \hline
\end{tabular}
\caption{A table showing the energy consumption of various bitcoin mining infrastructures. The first row (classical) is filled in using known real-world values. For the last three rows (Error-Code-Corrected (ECC) and non-ECC quantum miners, the first column is calculated from first principles, and the final two columns are calculated to keep the same ratio as the classical miner.}\label{tab:TAB2}
\end{table}

The relevance of the first three rows should be immediately clear. Even with a fairly severe error-correction overhead of a two-layer concatenated Shor code, a quantum miner still massively outperforms a classical miner, consuming $2.94279275 \times 10^{8}$ times less energy. 

Table~\ref{tab:TAB2} shows results under the assumption that both quantum and classical miners have similar energy inefficiencies with respect to their theoretical optimums. We believe that this is a fair assumption---yet a reasonable question to consider is what happens if the assumption does not hold.

A reasonable question to raise would be, what if this assumption does not? Table~\ref{tab:TAB1} summarises just how much worse the efficiency of a quantum miner (again, compared to its theoretical optimum) would have to be in order for it to completely lose its advantage over classical miners. For instance, for a two-layer ECC quantum miner, the inefficiency would have to be $5.0204\times 10^{11}$ higher, than a classical computer, for it to completely lose its quantum advantage. 

Next, let us consider a non-ECC quantum miner. As can be surmised from the table, the level of quantum advantage in this case is \emph{staggering}: a non-ECC quantum miner delivers $1.575753 \times 10^{21}$ energy advantage over a similarly capable classical miner. This translates to an average $2.4457216 \times 10^{-15}J$ energy cost, per block mined, compared a classical miner's 2.256kJ.

It is worth pausing on this for one moment. Non-ECC quantum devices are often discounted, in comparison to ECC quantum devices, for one reason and one reason only: lack of scalability. For a fully-fledged, scalable, universal quantum computer, fault-tolerance is a key requirement. To quote from the EPSRC National Quantum Technologies Programme's \emph{Technical Roadmap for Fault-Tolerant Quantum Computing:} ``qubits don’t necessarily need to be error-corrected. It might be possible to use physical qubits that are good enough to undergo thousands of operations before an error occurs. However, such a machine is not scalable, and could only perform short algorithms.''\cite{EPSRC}

A quantum miner is \emph{not}, and need not be, a scalable, universal quantum computer. A quantum miner need only perform a single task. Hence, it can be a highly specialised piece of kit, more akin to current specialised ASIC miners, rather than universal CPUs. The task they need to complete is (unlike most other applications of quantum computation) \emph{easy enough to be solvable classically.} And finally, not only is the problem solvable classically, it is routinely solved by classical devices \emph{in under ten minutes.}---the block time of Bitcoin. Depending on `clock speed' a quantum miner could complete its task within a few seconds, or even under a second (see the Methodology Section for details).

Finally, as we discuss further in the Conclusions, given the nature of cryptocurrency mining, a quantum miner may have a fairly large probability of failing its computation, and still be quite profitable. All this makes Bitcoin mining, and in general blockchain PoW-mining, an ideal candidate application for NISQ technologies.

\section{Methodology}\label{sec:MET}

Here, PoW will be considered to be a hash-based, specifically HashCash \cite{Back2002} implemented by the Bitcoin network. 

We have analysed three different quantum mining architectures, as mentioned above: a one-layer Shor-code error-corrected quantum miner, a two-layer Shor-code one, and a non-ECC quantum miner. On each of three architectures we run a quantum mining algorithm---essentially Grover's search algorithm on the appropriate search space---such that the quantum miner achieves  the same probability of mining a block per block cycle as a current classical device. The precise quantum algorithm is given in the Appendix.

In order to estimate the energy costs of our three analysed architectures, we proceed as follows. First, we calculated the Landauer limit of a classical miner. We then compare that to the actual energy costs of a current ASIC-based miner, in order to derive an efficiency ratio. In this case, we arrived at two different ratios. The first, $1:379$, is based on the manufacturers declared energy consumption per hour. The second is calculated from the CBECI-advertised energy costs\cite{Cambridge_Cbeci}, this derivation of the ASIC-miner energy costs caused us to arrive at a slightly different ratio of $1:1706$.

The classical miner we chose for our analysis is the Antminer S19 XP \cite{Bitmain}.This device provides 140 TeraHashes(TH)/s. This means that 84000 TH per 10 minute block time can be produced. 

The current bitcoin hash rate at time of writing is approximately 201 million TH/s (120.6 billion TH per block), we can therefore calculate that the Antminer ASIC device would represent approximately $0.000070\%$ of the current mining hash rate on Bitcoin. The current energy consumption per block is approximately 3.2267GJ and 0.000070\% of this is 2.25869kJ per block by our mining device. Therefore the efficiency ratio of our ideal device compared with our actual device is 1:1706. We will also consider a second ratio, \cite{Bitmain} states that the Antminer S19 XP has an energy consumption of 3.010kW (502W/block), this give us a ratio to the value given by the manufacturer of 1:379.

Using Landauer's principle we can easily calculate the minimum required energy $E$ as follows\cite{peterson2016experimental}:
\begin{equation}\label{eq:landauer}
E = k_BTln2\cdot B,
\end{equation} 
where  $k_B$ is Boltzman constant and $T$ is the temperature of the heat sink and $B$ is the number of bits erased. In the cases for both classical and quantum devices we will consider room temperature as the heat sink. 
Each block produces a hash value of 256 bits, and 8588 NAND gates are needed to produce each hash\cite{kim2008efficient}.
Each NAND gate requires 0.625bits worth to be erased. Hence, in total, 5368 bits are erased per hash function . We can therefore calculate that per block period if the classical device performs the 84000 TH then $4.72\times10^{20}$ bits of data are erased per block. We then feed these numbers back into Eq.~\ref{eq:landauer} to calculate that the minimum possible energy of an idealised classical device is approximately 1.324J per block (1.324 J). 

Next, for each of the analysed architectures we derive the Landauer limited energy costs of executing the mining algorithm on said architecture. We then multiply these costs by the above ratios to achieve our predicted energy costs. These results are displayed in table \ref{tab:TAB2}.

\begin{table}[ht]
\centering
\begin{tabular}{|p{3.0cm}|p{2.5cm}|p{2.5cm}|}
\hline
\textbf{Infrastructure} &
  \textbf{Landauer Theoretical Minimum (J)} &
  \textbf{Ratio to Equal Classical Actual} \\ \hline
Classical                      & 1.324   & 1:1706  \\ \hline
Non-ECC NISQ Miner         & $1.43 \times10^{-18}$ & $1:1.58 \times10^{21}$\\ \hline
1 Layer ECC Quantum Miner & $2.5 \times10^{-10}$  & $1:6.02\times10^{12}$ \\ \hline
2 Layer ECC Quantum Miner & $4.5 \times10^{-9}$  & $1:5.02\times10^{11}$ \\ \hline
\end{tabular}
\caption{This table shows the required efficiency ratio for various infrastructures to equal the energy consumed by the classical device per block cycle. For the the first row, the ratio is derived directly from the known energy costs of ASIC miners, and the theoretical minimum calculated here. For the quantum miners the ratios are calculated so that the energy costs of each miner equate that of the classical miner.
}\label{tab:TAB1} 
\end{table}

We begin with a non-ECC quantum device.
 During the execution of the quantum algorithm no data is erased. The only erasure occurs once the output, consisting of 512 qubits, is measured.  This is constant regardless of the mining difficulty parameter, and of the desired successful mining probability. Using again Eq.~\ref{eq:landauer} the minimal energy expenditure of a non-ECC quantum miner is $1.4336 \times10^{-18}J$. Using the first ratio calculated above of $1:379$ this quantum device would be expected to consume $5.4 \times 10^{-16}J$. With the ratio of $1:1706$ this would become $1.4 \times 10^{-15}J$ per block mined.

Next, we analyse an error-corrected quantum miner that uses a single layer of the Shor code. In this case, we must consider the success rate, or probability of successfully mining a block. We use same probability ($\approx 0.0000007$) as the chosen classical miner for finding a correct block per block cycle.

First, we calculate the total needed number of Grover's iterations needed to calculate PoW with a probability close to $1$. To do this we must find the target value of the PoW problem. This is calculated by:
$
   M = \frac{maxTarget}{Difficulty}
$, where $maxTarget$ and $Difficulty$ are two blockchain-dependant parameters that are tweaked over time, in order to maintain a constant block time, or average time it takes to add a block to the blockchain---which in the case of Bitcoin is 10 minutes.

The number of Grover's iterations can then be calculated:
$t = \sqrt{\frac{N}{M}}$,
where N is the size of the total search space. For Bitcoin, this this is currently $2^{256}$.

The total number of (error-corrected) gates the miner needs to run for each blocked mined is needed. This is given by:
   $E = t \times g$
where, $g$ is the number of gates per Grover's iteration. This number can easily be discerned from the pseudo-code / circuit of the quantum miner (see the Apendix/Supplemantal Material), giving a value of $1280$.

We must also consider the depth of the quantum circuit, since even qubits which are not currently undergoing any gates must be error-corrected. The depth of the circuit is given by 
$EC Steps = E\times d$,
 where $d$ is the number of qubits to be error corrected. 

 This can give us the overall equation: 

 \begin{equation}
     EC Steps = \sqrt{\frac{N}{(\frac{maxTarget}{Difficulty})}} \times g \times d 
 \end{equation}

This gives a total number of error-correction steps to be $1.1159814903 \times 10^{10}$

Given that every error-correction step in the Shor code requires the measurement of 12 qubits over two rounds of measurement, this means that a total of 
    $errasedBits = ECSteps \times c^n + q$
bits need to be erased, where c is the number of error correcting measurements per round of error correction, n is the number of layers of error correction and q is the number of qubits erased once the algorithm is completed. This comes to the totals of  of $1.339 \times 10^{11}$ and $1.607 \times 10^{12}$ bits needing to be erased for one layer and two layers respectively, during the run of the mining algorithm. 

Once again, using Eq.~\ref{eq:landauer} the minimal energy expenditure of a one-layer Shor-code ECC quantum miner is $3.7497 \times 10^{-10}J$. Using the first ratio calculated above of $1:379$ this quantum device would be expected to consume $1.4211 \times 10^{-7}J$. With the ratio of $1:1706$ this would become $6.39698 \times 10^{-7}J$ per block mined.

Using a similar calculation we can derive the optimal lower bound of the energy consumption of an ECC quantum miner using a two-layer Shor code. This comes to $4.4996 \times 10^{-9}J$. With the ratio of $1:1706$ this would become $7.67638 \times 10^{-6}J$ per block mined.

Finally, as mentioned in the introduction, a fair question regarding our methodology is whether it is reasonable to assume the same efficiency ratios for quantum miners as for ASIC-based miners. Rather than calculate energy costs for a slew of different efficiency ratios, one way to address this question is to calculate just how disparate the quantum efficiency would have to be compared to the classical efficiency, before quantum devices would stop have an energy costs advantage. A simple calculation shows that that the efficiency ratio for a 2-layer ECC quantum device would have to be $5.0204 \times 10^{11}$. These results are summarised in Table~\ref{tab:TAB1}.

\section{Conclusion}\label{sec:conc}

In short, we have shown that under the reasonable assumption that quantum computers will be similarly energy inefficient---with respect to their theoretical optimum---as classical computers,  the former will provide a very sizeable energy-cost advantage over the later. These projected energy savings would amount, over a year, to a total of 126.7TWH. To put this into perspective this is only slightly less than the total energy consumption of Sweden in 2020 \cite{IEA_Sweden}.

It is possible that quantum computers stray, even considerably, from these projections. However, we show that quantum computers would have to $1.1296 \times 10^{12}$ orders of magnitude less efficient (compared to their theoretical optimum) than classical computers, before they lose their inherent quantum energy-efficiency advantage. In table \ref{tab:TAB2} we show further results.

While some newer cryptocurrencies have moved, or are in the process of moving, away from Proof-of-Work (in part due to energy costs of keep the blockchain running), Bitcoin, the biggest cryptocurrency, by far, in terms of both market capitalization/value and number of users, is currently using Proof-of-Work, and there are no feasible plans to change that.

These potential energy saving represent an enormous technological opportunity for two major reasons. The first is ecological.  We present here a feasible technological avenue to reduce our energy footprint considerably. 

The second is the impact on quantum technologies. Applications of quantum technologies, and in particular applications of NISQ technology, has been a major area of interest in recent years. In particular, NISQ technologies are currently not considered to have very impactful applications; to quote J. Preskill,
 one generally \emph{` shouldn’t expect
NISQ to change the world by itself'}\cite{preskill2018quantum}.

Definitions of NISQ often vary, but the term is generally used to refer to quantum computers with no more than a few hundred qubits. At $512$ qubits, a quantum bitcoin miner clocks in a bit over that, but not by much. 
Quantum cryptocurrency mining has many other characteristics that make it particularly suitable to a NISQ implementation.

First, these devices are not meant to solve intractable problems, but rather to solve \emph{very} tractable problems (solvable within 10 minutes on a classical device). The quantum device merely needs to achieve the same goal more \emph{efficiently.} Given this, it is feasible that a NISQ miner need only run a few seconds (or even factions of) before it can be safely reset. This makes usual considerations of quantum devices of heat, entropy, $t_1$ and $t_2$ relaxation times, much more tractable. 

A second consideration is that the computation success-rate need not be high at all. A classical Bitcoin miner is profitable with only a success-rate of about $0.000070\%$. Indeed, once the hardware is fixed, a classical miner cannot improve its successful mining probability in any way. Hence, a NISQ miner could be be very noisy, and still compete favourably against classical miners.
A NISQ miner could further potentially fine-tune its success probability (trading  runtime) by adjusting the number of Grover iterations. 

Obviously, both points above also mean that fault-tolerance circuitry can be forgone entirely, without overly hindering the quantum miner.

The above characteristics make cryptocurrency mining a very reasonable application of NISQ (or NISQ-adjacent) devices that do indeed have a world-changing potential.

\section{Acknowledgements}

The authors would like to acknowledge funding through the  Quantum Communications Hub (EP/T001011/1), and from the Casper Association Academic Grants Program.


\bibliography{bibliography}
\bibliographystyle{unsrtnat}

\pagebreak
\appendix

\section{Quantum Mining Algorithm}

Algorithm \ref{prot:2} lays out the pseudo-code for a quantum Proof-of-Work miner. The goal of the miner is to take a set of transaction, create a new block with them, and then find a nonce value such that the value calculated by a given secure hash function, operating on the new block and nonce, is below a certain target value. The search space for the desired nonce is $2^{256}$
On a quantum miner, the search for a particular nonce is done using Grover's quantum search algorithm. Given the size of the search-space, we get total of 1280 gate operations per Grover's iteration. 
The total number of Grover's iterations will depend on the required or desired success probability.

\begin{algorithm}
\caption{Quantum Proof of Work}\label{prot:2}

\begin{description}
\item[Miner's Block Input] 
\begin{enumerate}
\item Version Number
\item Hash of Previous Block
\item Merkle Root Hash
\item Time Stamp
\item Difficulty Target $D$
\item Nonce $i$
\item Block Size
\item Transaction Counter
\item Transactions $T_1 \dots T_n$
\end{enumerate}
\item[Output] Block $B$ such that $\mathfrak{H}(B) \leq B_{target} $
\item[Steps]
\begin{enumerate}
\item Create block $B$ from Miner's Block input.
\item Set  $B_{target}$ through the calculation $2^{256} - (D \cdot 2^{32})$
\item Set up Oracle $\mathfrak{O}$ that marks solutions where $\mathfrak{H}(B) \leq B_{target} $
\item Set up $n$ qubits for the $n$ bit search space(generally $2^{256}$ or 32-Byte value for $i$)
\item Initialize the system to a uniform superposition of all possible nonce values by applying Hadamard transform $H^{\otimes n}$
\item Perform Grover's algorithm iterations, until the probability of finding a valid nonce is approximately 1. Each iteration consists of:
\begin{enumerate}
\item Apply $\mathfrak{O}$ over the search space 
\item Apply Hadamard transform $H^{\otimes n}$
\item Perform phase shift such that $\ket{0} \rightarrow \ket{0}$,  $\ket{x} \rightarrow -\ket{x}$ for $x > 0$
\item Apply Hadamard transform $H^{\otimes n}$
\end{enumerate}
\end{enumerate}
\end{description}
\end{algorithm}

\end{document}